\documentclass[
    ,final            
  ,numberedheadings 
  ,sort&compress                
  ]
  {aipproc}

\usepackage{aas_macros}
\usepackage{floatfig,epsfig}

\layoutstyle{6x9}

\begin{document}
\initfloatingfigs

\title{Galactic Binaries as Sources of Gravitational Waves}

\author{G. Nelemans}{
  address={Institute of Astronomy, Madingley Road, Cambridge CB3 0HA, UK}
}

\begin{abstract}
  I review the expected Galactic sources of gravitational waves,
  concentrating on the low-frequency domain and summarise the current
  observational and theoretical knowledge we have. A model for the
  Galactic population of close binaries, which is tested against the
  observations is used to predict the expected signal for the future
  space based gravitational wave detector LISA. With a simple model
  for the electro-magnetic emission from the same binaries I argue
  that a fair number of the LISA systems have electro-magnetic
  counterparts, which can be used to improve in particular the
  distance and mass measurements of these systems by LISA.
  Furthermore, LISA will enable us to test some aspects of the theory
  of binary evolution that are very difficult to asses in different
  ways.
\end{abstract}

\maketitle


\section{Introduction: Galactic sources of GWR}

As for all astrophysical phenomena, if it is present in our own Galaxy
it often can be studied in most detail, simply because of the
proximity of the sources. This also holds in some sense for
gravitational wave phenomena. Many Galactic sources are, or could be
(strong) gravitational wave radiation (GWR) sources. In particular
binary stars are obvious sources as has been realised long ago
\citep[e.g.][]{mir65}.

The amplitude of the gravitational wave signal increases with the
chirp mass of the binary and its gravitational wave frequency
\citep[e.g.][]{eis87}, which means that more massive, equal mass,
short period binaries are the most promising sources. The frequency
range covered by the current and planned detectors is limited to to
frequencies between about 0.1 mHz and 1 Hz for the space based
detector LISA and about 50 to 5000 Hz for ground based detectors. For
binary objects this translates to orbital periods between 5.5 hr and 2
seconds and 40 and 0.4 milliseconds for space and ground based
detectors respectively. These short periods imply small separations
(or very high masses) through Kepler's law, thus for stellar mass
objects this means that we have to concentrate on compact stars, in
particular helium stars, white dwarfs, neutron stars and black holes.

\begin{figure}
  \includegraphics[width=\textwidth]{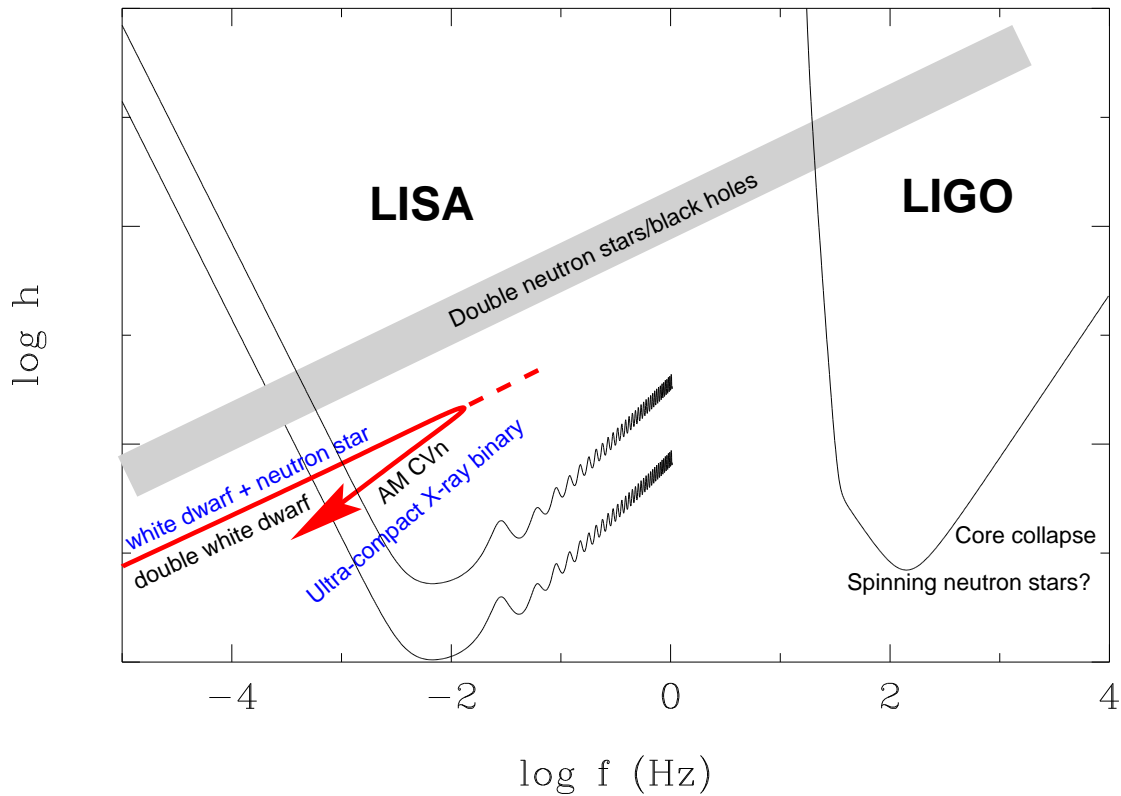}
  \caption{
    Schematic picture of the evolution of compact binaries in
    frequency -- gravitational wave amplitude space. Plotted are the
    expected sensitivities of LISA and LIGO, and the evolution of
    different types of binaries, discussed in the text. The big arrow
    shows the evolution of binaries that start stable mass transfer,
    while the dashed extension in the middle of the plot represents
    binaries which have more massive white dwarfs, evolving to shorter
    periods, where they coalesce. }
  \label{fig:intro}
\end{figure}

The most important classes of binary stars as sources of gravitational
waves are: helium star -- white dwarf/neutron star binaries, double
white dwarfs, white dwarf -- neutron star/black hole binaries, and
double neutron star/black hole binaries (see Fig.~\ref{fig:intro}). If
these objects form in binaries with orbital periods below $\sim$10 hr,
the angular momentum losses due to GWR will make them spiral together
within a Hubble time, until the separations are small enough that mass
transfer starts. Helium stars start mass transfer typically at periods
of 30 -- 60 min (GWR frequencies of 1 -- 2 mHz), but evolve to shorter
periods, before they reach a period minimum around 10 min
\citep[e.g.][]{skh86,tf89}. Double white dwarfs and white dwarf --
neutron star/black hole binaries start mass transfer at orbital
periods of a few minutes (GWR frequencies around 20 mHz). The more
massive the white dwarf in these binaries, the smaller it is and thus
the shorter the period at which mass transfer starts.

Both the helium star as the white dwarf binaries might start
\emph{stable} mass transfer to their white dwarfs, or neutron
star/black hole companion, causing these binaries to evolve back to
longer periods (see Fig.~\ref{fig:intro}). Such sort-period
mass-transferring objects are observed and are called AM CVn systems
and ultra-compact X-ray binaries (UCXB's) in the case of white dwarf
and neutron star/black hole accretors respectively.

Finally binaries in which both components are neutron stars or black
holes will continue decreasing their orbital until they reach stunning
orbital periods of about 1 ms (GWR frequency about 2 kHz, see
Fig.~\ref{fig:intro}). These frequency ranges are also the ranges
where non-binary Galactic sources, like rapidly rotating neutron stars
and and the rapidly rotating cores of collapsing stars in supernova
explosions will emit GWR \citep[e.g.][]{aks98,bil98,rmr98}. For a
discussion of these high frequency sources, I refer to contributions
of Mezzacappa, Fryer, Heyl, Bulik in this volume. For the remainder of
this article I will concentrate on Galactic sources of
\emph{low-frequency} GWR.

\pagebreak[4]
As our knowledge of these sources stems from a combination of
observational facts and model extrapolations of these observations, I
will first discuss the observations we have of the short period binary
populations (Sect.~\ref{obs}), before discussing a model for the
Galactic short period binaries (Sect.~\ref{model}) and presenting the
expected low-frequency signals that can be detected by LISA
(Sect.~\ref{LISA}). I will then discuss the importance of
complementary electro-magnetic observations (Sect.~\ref{EM}) and the
scope for ``Galactic GWR astronomy'', i.e. testing our models and
understanding (in this case of close binaries) with GWR measurements
(Sect.~\ref{GWR_astro}).

\section{Summary of current observational knowledge}\label{obs}

\addtocounter{figure}{1}
\begin{floatingfigure}{6cm}
\mbox{\includegraphics[angle=-90,scale=0.5,clip]{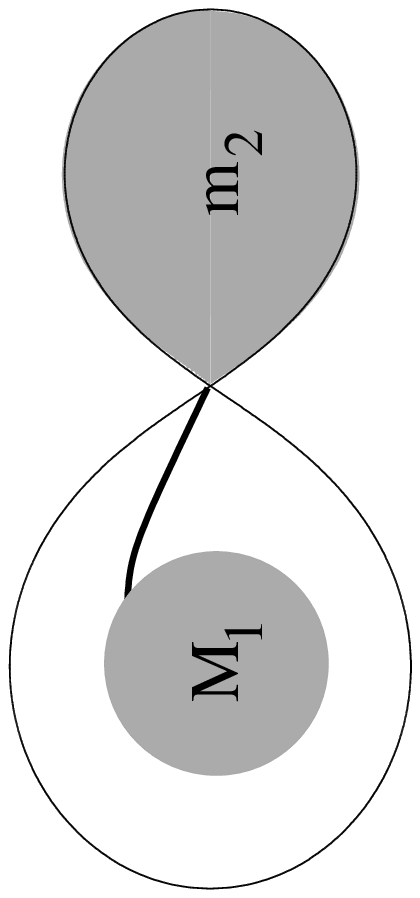}}\\
\vspace*{0.2cm}

\noindent{\fontsize{10}{11.5}\selectfont \textbf{FIGURE 2.} Schematic picture of the direct impact phase of accretion
  between two white dwarfs.}
\label{fig:DI}
\end{floatingfigure}
\par
For a more comprehensive discussion of the observed short period
binaries that are relevant for LISA see \citet{ver97,vn01}. In the
last few years there have been quite some new discoveries of compact
binaries, which will enable us to test the theoretic models in some
more detail than was previously possible. I summarize these below.

\vspace*{0.7cm}
\begin{description}
\item[\textbf{Double white dwarfs}] Up to the middle of the 1990s,
  about 200 white dwarf were checked for binarity, yielding 14 close
  double white dwarfs \citep[see for a review][]{mar00}. The SPY
  project \citep{ncd+01} is a survey on the ESO VLT 8m telescope,
  using the UVES spectrograph to check $\sim$1000 white dwarfs for
  radial velocity variations indicating binarity. For each white
  dwarf, two spectra with high spectral resolution and high
  signal-to-noise are taken and compared. It will be, by far, the
  largest (fairly homogeneous) sample of white dwarf checked for
  binarity and is expected to discover $\sim$150 - 200 new close
  binary white dwarfs. The current status is that 577 white dwarfs
  have been analysed and 123 are close binaries, 14 with an M dwarf
  companion, the rest with a white dwarf companion. For about ten
  systems follow-up observations to determine system parameters like
  orbital periods, masses etc. have been completed, yielding orbital
  periods between 0.3 and 5 days
  \citep[e.g.][]{neh+01,nkn02,knh+02}. In addition to the SPY
  discoveries, a double white dwarf with a relatively long period of
  30 days was found by \citet{mbm+02}.

\item[\textbf{AM CVn systems}] Very excitingly, there have been three
  AM CVn candidates found with extremely short periods: RX
  J1914.4+2456 \citep[V407 Vul,][with a possible period of 9.5
  minutes]{chm+98}, KUV 01584-0939 \citep[ES Cet,][with a period of 10.3
  minutes]{ww02} and RX J0806.3+1527 \citep[][with a possible period
  of 5.3 minutes]{rhc02,ihc+02}. Not only does this bring the number
  of known AM CVn systems from 8 to 11, but it shows that the shortest
  period AM CVn systems, which are much less numerous than their
  longer period descendants because they evolve quickly, can be found.
  The two shortest period systems might also be in a so called
  ``direct impact'' phase of accretion, in which the accretion stream
  directly hits the surface of the accreting star rather than forming
  a disc \citep[see Fig.~\ref{fig:DI}]{npv+00,ms02}.
  Based on a simple modelling of the brightness of the systems we found
  that most of the systems in a magnitude limited sample would fall
  between roughly 20 and 40 minutes \citep{npv+00}. Indeed, two of the
  new candidates are found by their X-ray emission, different from the
  ``normal'' AM CVns that are found in the optical. The two shortest
  period systems await spectroscopic confirmation of their periods to
  be orbital, and some models suggest they are not AM CVns but either
  detached double white dwarfs \citep{wcr02} or longer period systems
  \citep{nhw02}. Recently another suspected supernova (SN2003aw)
  turned out to be an AM CVn system undergoing an outburst
  \citep{cf03}.
  
\item[\textbf{Ultra-compact X-ray binaries}] Another recent
  development, worth mentioning is the discovery of three new
  ultra-compact X-ray binaries (where with ultra-compact I mean
  orbital period below the period minimum for a main sequence mass
  donor ($\sim$60 min), bringing the total to 7 of which 2 reside in a
  globular cluster. In all systems the neutron star turned out to be
  an accreting millisecond X-ray pulsar: XTE J1751-305
  \citep{ms02a,ms02b,mss+02}, XTE J0929-314
  \citep{rss02,gmr+02,gcm+02} and XTE J1807-294 \citep{mjs03}. The
  systems have orbital periods of 42.4, 43.6 and 40.1 mins
  respectively.
  
  There are another six candidates, based on similarities with known
  ultra-compact X-ray binaries in either X-ray and/or optical spectra,
  or optical brightness \citep[e.g.][]{jpc00,dma00}.
\end{description}

\section{A model for the Galactic population}\label{model}

To be able to predict the number of low-frequency GWR sources we have
to construct a model for these binaries in the Galaxy, as the
observations are too incomplete to estimate the total number of
systems present in the Galaxy.  As our knowledge of stellar and in
particular binary evolution is rather limited, we have to construct
this a model in such a way that we can use the observations to test
some of the basic assumptions that go into the model. The basic
ingredients of a Galactic binary population model are

\begin{itemize}

\item A description of stellar and binary evolution:

\begin{itemize}
  
\item For single stars, we need their mass, radius, luminosity and
  core mass, as function of initial mass and time (and
  metallicity).
\item For binary stars a recipe for effects of stellar wind mass loss,
  mass transfer, supernova explosions etc on the orbit of the binary.
\end{itemize}

\item Initial parameter distributions

\begin{itemize}
\item Initial primary mass distribution, according to some initial
  mass function, the mass ratio distribution of the zero-age main
  sequence binaries, an initial separation distribution and initial
  eccentricity distribution of the binaries.
\end{itemize}

\item Normalization and space distribution

\begin{itemize}
\item Star formation history of the Galaxy
\item The initial binary fraction of objects in the Galaxy.
\item Galactic distribution, of stars (as function of time)
\end{itemize}

\end{itemize}

We use the binary star population synthesis code developed by
\citet{pv96}, with some new ingredients like the initial mass function
\citep[for which we now use][]{ktg93}, and a self consistent model for
the star formation rate as function of time and position in the
Galaxy, based on \citet{bp99}. For further details on the ingredients
of the model presented here, we refer to \citet{nyp03}. In
Table~\ref{tab:rates} I show the resulting birth and merger rates in
the Galaxy for binaries containing compact objects.  An order of
magnitude estimate in the uncertainty in the quoted birth and merger
rates is given in the table. For the double white dwarfs this number
comes from the range we get in our own models
\citep[see][]{nyp+00,nyp03} plus the estimate of the fraction of white
dwarfs that are close binaries \citep{mm99}. For AM CVn systems the
estimate is based on \citet{npv+00}. For the neutron star binaries the
estimates are based on the range of birth rates published by
\citet{py98,bkb02}.

\begin{table}
\caption{Galactic merger rates and birth rates for binaries containing compact
  objects, with an (order of magnitude) estimate of the uncertainty and 
  the number of systems that can be resolved by LISA (see
  Sect.~\ref{model} and \ref{LISA}).}
\label{tab:rates}
\begin{tabular}{lrrrr}\hline
Type             & \tablehead{1}{r}{b}{birth rate}   & \tablehead{1}{r}{b}{merger rate}   &  \tablehead{1}{r}{b}{uncertainty} &  \tablehead{1}{r}{b}{resolved systems} \\
                 & \multicolumn{1}{c}{(yr$^{-1}$)}& \multicolumn{1}{c}{(yr$^{-1}$)} & factor  & \\ \hline 
(wd, wd)         & 2.0 $\! \times \!$ 10$^{-2}$   & 8.3 $\! \times \!$ 10$^{-3}$    & 5 & 10658   \\
AM CVn           & 1.3 $\! \times \!$ 10$^{-3}$   &                                 & 50 & 9831   \\
UCXB             & 1.1 $\! \times \!$ 10$^{-5}$   &                                 & 20 & 22       \\
 (ns, wd)        & 6.8 $\! \times \!$ 10$^{-5}$   & 3.8 $\! \times \!$ 10$^{-5}$    & 20 & 8      \\
 (ns, ns)        & 5.2 $\! \times \!$ 10$^{-5}$   & 2.5 $\! \times \!$ 10$^{-5}$    & 50 & 7      \\
 (bh, wd)        & 7.2 $\! \times \!$ 10$^{-5}$   & 2.6 $\! \times \!$ 10$^{-6}$    & 50 &0      \\
 (bh, ns)        & 3.5 $\! \times \!$ 10$^{-5}$   & 1.0 $\! \times \!$ 10$^{-5}$    & 50 &0      \\ 
 (bh, bh)        & 1.7 $\! \times \!$ 10$^{-4}$   & 5.1 $\! \times \!$ 10$^{-6}$    & 50 &0      \\ \hline
\end{tabular}
\end{table}

\section{Expected results from LISA}\label{LISA}

We now use the Galactic model described in the previous section to
calculate the expected signals that are detectable by LISA as we did
in \citet{nyp01}, see also \citet{eis87,hbw90,wh98,pp98,hb00}.  In
Table~\ref{tab:rates} we list the number of systems that LISA will be
able to resolve individually in frequency. 

\begin{figure}
  \includegraphics[height=\textwidth,angle=-90]{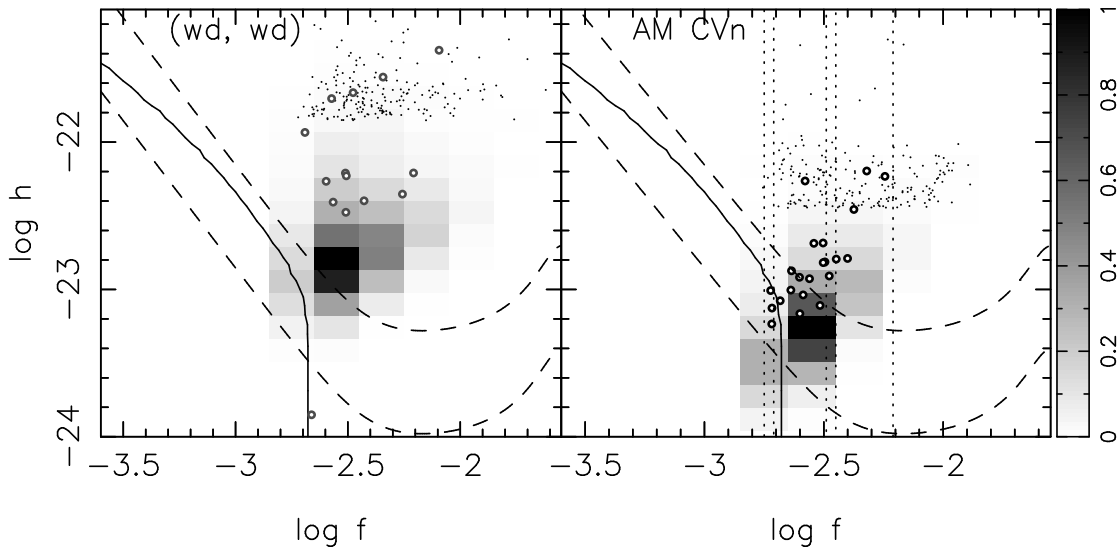}
  \caption{GWR wave amplitude $h$ as function of the GWR frequency $f$
    for the expected low-frequency Galactic binaries, which are
    expected to be detectable by LISA. The left panel shows the
    (10658) double white dwarf systems as the grey shade, with the 200
    strongest sources as points, to increase their visibility. The
    right panel shows the (9831) resolved AM CVn systems that are
    expected, again showing the 200 strongest sources as points. Over
    plotted with the large symbols are the neutron star binaries in
    the left panel and the ultra-compact X-ray sources in the left
    panel. The average double white dwarf background is plotted as the
    solid line, while the dashed curves show the LISA sensitivity for
    a integration time of 1 yr giving S/N of 1 and 5 respectively
    \citep{lhh00}.}
\label{fig:fh}
\end{figure}

Fig.~\ref{fig:fh} shows the expected signals detected by LISA in
frequency space, i.e. neglecting any complications from the fact that
LISA is actually moving in space. At the lowest frequencies the vast
number of double white dwarfs in the Galaxy form an unresolved noise
background. The average of the background is plotted as the solid
line. Only at frequencies above $\sim$2 mHz the background disappears
and individual systems can be resolved. The distribution of GWR wave
amplitude vs.  frequency of the resolved double white dwarfs and AM
CVn systems is plotted as the grey shades, while the 200 strongest
sources, plus the resolved detached neutron stars binaries and
ultra-compact X-ray binaries are shown individually.

The vast majority of the expected resolved sources are double white
dwarfs and AM CVn systems, with only a handful of neutron star
binaries and ultra-compact X-ray sources. See \citet{nyp03} for more
details.

A few remarks must be made about these calculations. The number of
resolvable systems is determined in a simple way, just evaluating the
number of systems present in each frequency bin, with fixed width,
$\Delta f = 1/T$, where $T$ is the integration time, for which we use
1 yr. That means that the exact response of the detector to the
combined signals isn't taken into account, which on the one hand might
complicate the actual detection of the sources.  We also did not
consider the possibility that sources could be detectable in the
frequency range where the double white dwarf background noise
dominates, either because of the large amplitude or possibly their
position on the sky. For a discussion of the number of resolved
systems for which the frequency change can be measured see e.g.
\citet{wh98,nyp01,nyp03}.

\section{Complementary electro-magnetic observations}\label{EM}

\addtocounter{figure}{1}
\begin{figure}
\begin{minipage}[b]{0.7\textwidth}
\includegraphics[height=12cm,clip]{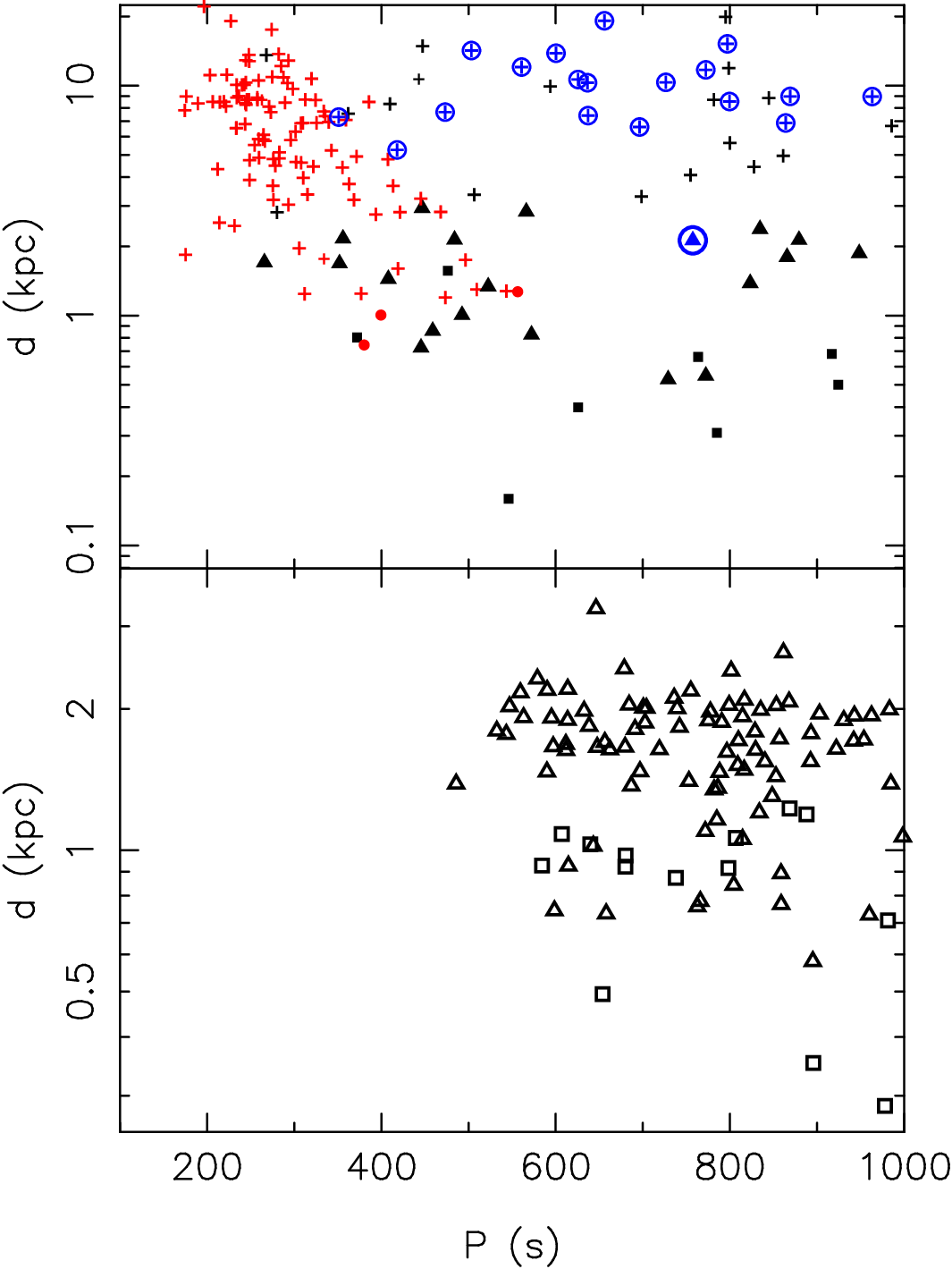}
\end{minipage}
\begin{minipage}[b]{0.28\textwidth}
\noindent{\fontsize{10}{11.5}\selectfont \textbf{FIGURE 4.} 
  Period vs distance distribution of resolved systems detectable in
  the X-ray and/or optical band. \textbf{Top panel}: Systems
  detectable only in X-rays (pluses), and systems that are detectable
  in the optical and X-ray band (filled circles: direct impact systems
  with a bright donor, filled triangles: systems with an optically
  bright disc, filled squares: systems with bright disc + donor). The
  ultra-compact X-ray binaries are the circled symbols. \textbf{Bottom
    panel}: Systems only detectable in the optical band (circles:
  direct impact systems with a bright donor, triangles: bright discs,
  squares: bright disc + donor). From \citet{nyp03}.}  \vspace*{1.2cm}
\end{minipage}
\end{figure}

As briefly mentioned before, LISA will in fact measure the signals
from the binaries convolved with the changing detector response due to
its orbit in space. That means that the position of the source can be
reconstructed \citep{cut98}, but it also means that the measured
signal depends not only on the binary properties (period, and masses)
and the distance, but also on the position in the sky (and the
relative contributions of the two GWR polarization signals depend on
the inclination of the binary). An interesting possibility is using
electro-magnetic observations to determine some of the parameters
(like position on the sky and orbital period) of the systems that can
be resolved by LISA, in order to use the GWR data to determine other
parameters that are difficult to obtain otherwise (masses,
inclinations and, if the frequency change of source can be measure,
distance) to higher accuracy

We therefore use very simple models for the optical and X-ray emission
from the mass transferring systems (AM CVn systems and ultra-compact
X-ray binaries) to estimate the number of resolved system in our
Galactic model that can also be detected with optical or X-ray
detectors. The results are summarised in Fig.~4, where we plot the
subset of the resolved systems that can also be observed with current
X-ray (top panel) and optical (bottom panel) detectors. In total we
expect some 330 of the $\sim$10000 resolved AM CVn systems and almost
all resolved ultra-compact X-ray sources to be detectable in
electro-magnetic radiation. However, this estimate is based on the
capabilities of current optical and X-ray instruments, so by the time
LISA will be operational, this number will be higher.

These simple models only include X-ray emission from the direct impact
spot and the boundary layer, while the optical emission is either from
a single temperature disc, or from the donor star. We do not include
any irradiation or (compressional) heating of the accreting object.
For the interstellar absorption we use a very simple, symmetric,
smooth model. For more details, see \citet{nyp03}.

\section{Gravitational wave astronomy: testing the models with LISA}\label{GWR_astro}

\begin{figure}
\includegraphics[height=\textwidth,angle=-90]{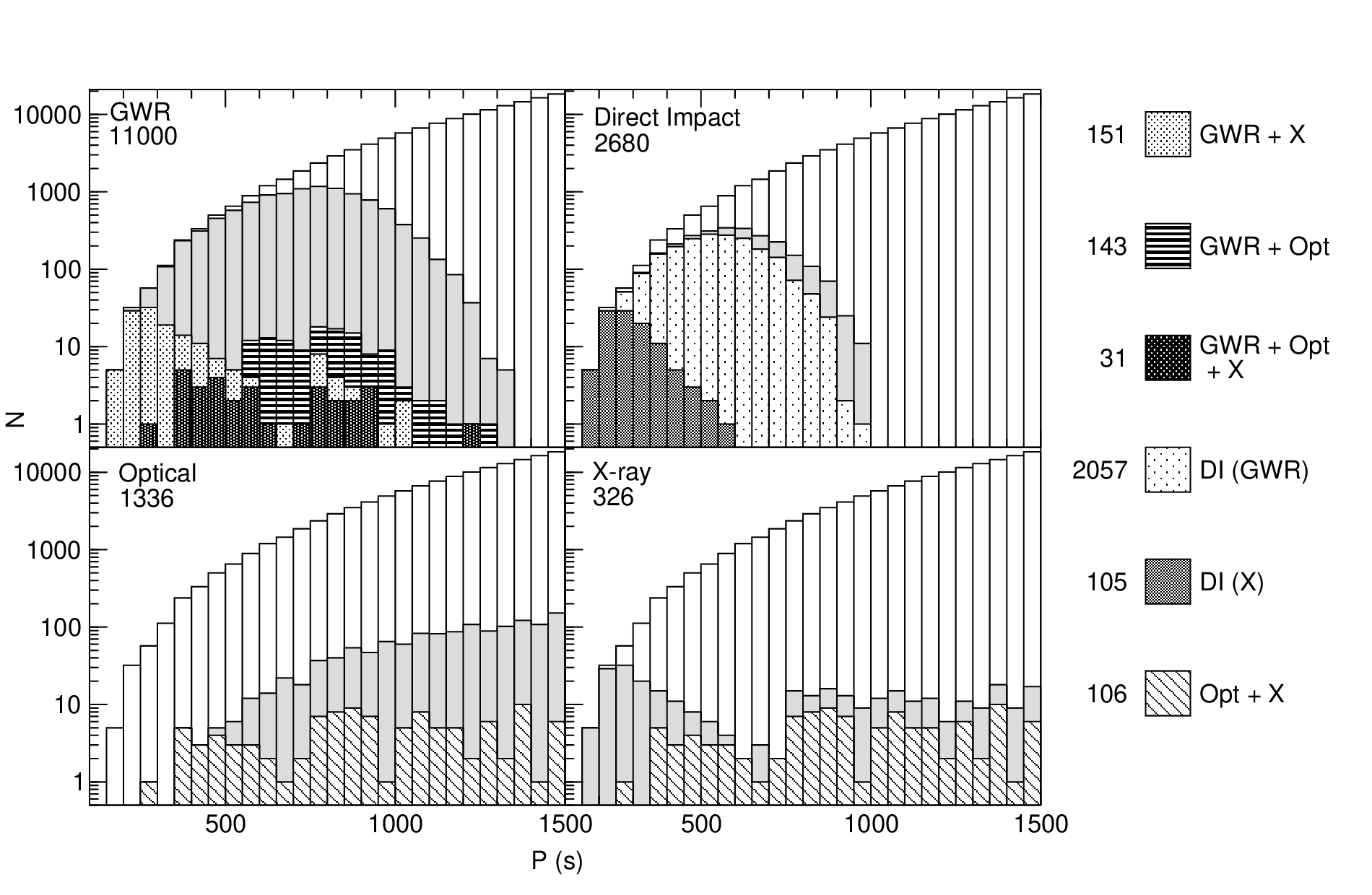}
\caption{Histograms of the population of short-period AM~CVn
  systems, subdivided in different types. In the top left panel we
  show the systems that can be resolved by \textit{LISA} in grey and
  subdivide these in the ones that have optical counterparts (GWR +
  Opt), X-ray counterparts (GWR + X) and both (GWR + Opt + X).
  Similarly the top right panel shows the population that is in the
  direct impact phase of accretion in grey and we subdivide that
  population in GWR and X-ray sources. The bottom two panels show
  (again in grey) the populations that are detectable in the optical
  band (left panel) and the X-ray band (right panel). In both lower
  panels the distribution of sources detectable both in optical and
  x-ray band is shown by the hatched region (Opt + X), from
  \citet{nyp03}}
\label{fig:hist_all}
\end{figure}

As shown in the previous sections, LISA will be able to discover a
very large number of short-period binaries in the Galaxy and will be
especially sensitive to the shortest periods. These periods are
exactly the ones that are most difficult to probe with classical
techniques. The double white dwarfs with periods below half an hour
are difficult to find, because the integration times used in
spectroscopic binary searches are typically of the order of 10
minutes, resulting in strong orbital smearing of the absorption lines.
Both for the detached systems and for the mass transferring systems
the time scale on which the period changes increases with decreasing
orbital period, so the number of systems at short periods is just much
smaller than at longer periods \citep[e.g. Fig.~2 of][]{nyp01}.

However, since LISA is sensitive mainly to the short period systems,
it will provide an enormous amount of information about the shortest
period range. In particular the phase directly before, and directly
after the onset of mass transfer has a number of question that are
almost impossible to answer with optical or X-ray observations. Just
before the mass transfer starts, the stars are so close together that
the usual assumption, that the angular momentum in the two stars and
the (tidal) interaction between the two stars can be neglected,
probably doesn't hold anymore. In that case the period change reflects
the combination of GWR losses and the other effects
\citep[e.g.][]{wh98}. 

At the onset of mass transfer almost all AM CVn systems will be in the
already mentioned direct impact phase.  The details of the evolution
of the binaries, and indeed the stability of the mass transfer in this
phase are very uncertain so that observations of these systems have to
provide the necessary information to understand, or at least probe
this phase \citep[e.g.][]{npv+00,mns02}.  To summarize our results we
show in Fig.~\ref{fig:hist_all} histograms of the different
subpopulations of short-period AM CVn systems \citep[see][for more
details]{nyp03}.

A second aspect of the large number of detectable systems is that, if
for enough systems the frequency change and thus the distance can be
measured, the short-period binaries can be used as tracers of Galactic
structure. In particular the mass distribution in the inner regions of
the Galaxy, which is difficult to observe because of interstellar
absorption, is a promising area of investigation.


\section{Conclusions}

I discussed our current knowledge of compact binaries in the Galaxy
and showed that there are still large uncertainties. With sometimes
rapidly increasing samples of observed systems some of the
uncertainties in the models can be addressed. However some questions,
especially related to the shortest period binaries are very difficult
to answer. The best current models predict a very large number of (in
particular) double white dwarf and AM CVn binaries that can be resolved
by LISA. As these are predominantly very short-period systems this
will provide invaluable information on some of the crucial open
questions. I argued that complementary optical, X-ray and infra-red
observations might be useful in constraining the parameters of the
resolved binaries, although many will not be detectable with
electro-magnetic detectors.


\begin{theacknowledgments}
  I would like to thank Lev Yungelson, Simon Portegies Zwart, Frank
  Verbunt and Sterl Phinney for help and discussions and PPARC for
  financial support.
\end{theacknowledgments}


\bibliographystyle{aipproc}   


\bibliography{journals,binaries}

\IfFileExists{\jobname.bbl}{}
 {\typeout{}
  \typeout{******************************************}
  \typeout{** Please run "bibtex \jobname" to optain}
  \typeout{** the bibliography and then re-run LaTeX}
  \typeout{** twice to fix the references!}
  \typeout{******************************************}
  \typeout{}
 }

\end{document}